\documentclass[letterpaper,twocolumn,10pt]{article}
\usepackage{usenix-2020-09}

\usepackage{tikz}
\usepackage{amsmath}
\usepackage{subfiles}

\usepackage{times,url,color,soul,xspace,enumitem,textcomp,booktabs}
\usepackage{graphicx}
\usepackage{caption}
\usepackage{subcaption}
\usepackage{comment}
\usepackage{xspace}
\usepackage{tikz}
\usepackage{amsmath}
\usepackage[inline,draft,nomargin,index]{fixme}

\hypersetup{breaklinks=true}

\usepackage{cleveref}
\crefformat{section}{\S#2#1#3} % see manual of cleveref, section 8.2.1
\crefformat{subsection}{\S#2#1#3}
\crefformat{subsubsection}{\S#2#1#3}

\newcommand{\Sec}[1]{\S{\ref{#1}}}

\FXRegisterAuthor{msb}{amsb}{\textcolor{red}{Msb}}

\FXRegisterAuthor{atr}{aatr}{\textcolor{blue}{Atr}}

\newcommand{\archName}{Hyperion}

\begin{document}

\author{
{\rm Marco Spaziani Brunella}\\
University of Rome Tor Vergata, Axbryd
\and
{\rm Marco Bonola}\\
CNIT/Axbryd
% copy the following lines to add more authors
 \and
 {\rm Animesh Trivedi}\\
VU, Amsterdam 
} % end author

\title{\archName: A Case for Unified, Self-Hosting, Zero-CPU Data-Processing Units (DPUs)}

\maketitle

\begin{abstract}
Since the inception of computing, we have been reliant on CPU-powered architectures. However, today this reliance 
is challenged by manufacturing limitations (CMOS scaling), performance expectations (stalled clocks, Turing tax), and
security concerns (microarchitectural attacks). To re-imagine our computing architecture, in this work we take a more
radical but pragmatic approach and propose to eliminate the CPU with its design baggage, and integrate three primary 
pillars of computing, i.e., networking, storage, and computing, into a single, self-hosting, unified CPU-free Data 
Processing Unit (DPU) called \archName. 
In this paper, we present the case for \archName{}, its design choices, initial work-in-progress details,  
and seek feedback from the systems community. 
\end{abstract}

\maketitle

\sloppy

\section{Introduction}
Since the inception of computing, we have been designing and building computing systems around the CPU as the 
primary workhorse. This primary architecture has served us well. However, as the gains from Moore’s and 
Dennard’s scaling start to diminish, researchers have started to look beyond the CPU-centric designs to 
accelerators and domain-specific computing devices such as GPUs~\cite{2011-sosp-ptask,2015-nsdi-gpu-packet-processing,2012-fast-shredder}, 
FPGAs~\cite{2014-isca-catapult,2020-osdi-os-on-fpga}, TPUs~\cite{2017-isca-tpu}, programmable-storage~\cite{2014-osdi-willow,2019-atc-insider,2021-atc-fcsv-virt-sto-fpga}, 
and SmartNICs~\cite{2019-atc-nica,2020-asplos-lynx}. The use of domain-specific computing devices in wide-spread 
mainstream computing is heralded as the \textit{Golden Age of Computer Architecture} 
by \mbox{Hennessy} and \mbox{Patterson} in their 2017 Turing Award lecture~\cite{2019-cacm-golden-age}.

\begin{figure}[t!]
\centering
\includegraphics[width=0.5\textwidth]{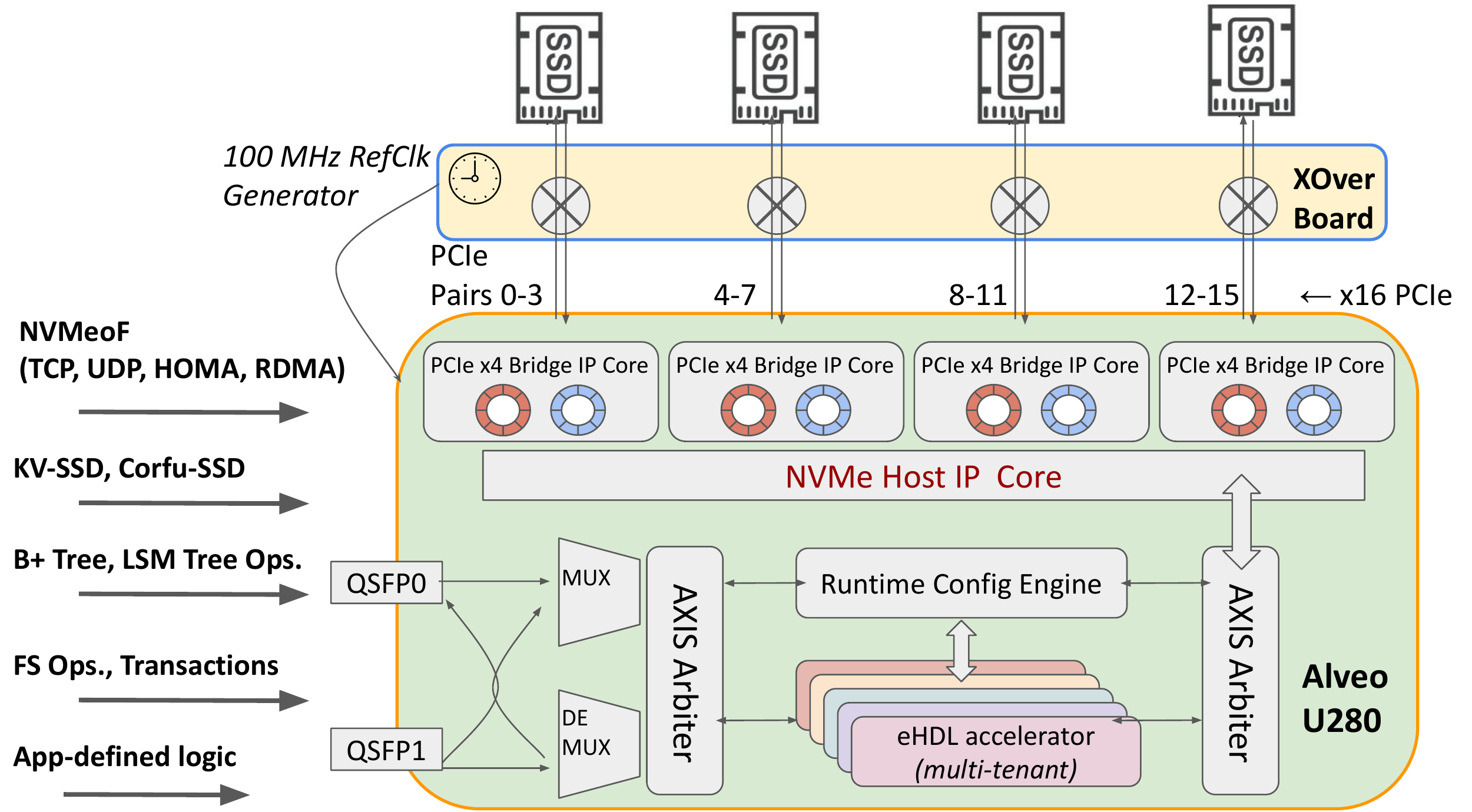}
\caption{\archName{} design with a unified network, FPGA, and NVMe SSDs.}\label{fig:arch}
\end{figure}

\begin{table*}[t!]
\small
\centering
\begin{tabular}{l p{0.75\textwidth} }
\toprule
\textbf{What} & \textbf{Examples} 
\\
\midrule
Network + Accelerator & SmartNICs~\cite{2022-bluefield,2021-asplos-nic-offloading}, AcclNet~\cite{2018-nsdi-azure}, hXDP~\cite{2020-osdi-hxdp} \\
Network + GPU & GPUDirect~\cite{2022-nvidia-gpudirect}, GPUNet~\cite{2014-osdi-gpunet} \\  
Storage + GPU & Donard~\cite{2022-donard}, SPIN~\cite{2017-tos-spin}, GPUfs~\cite{2013-asplos-gpufs}, GPUDirect~\cite{2019-gpudirect-storage}, nvidia BAM~\cite{2022-arvix-bam}\\ 
Network + Storage & iSCSI, NVMoF (offload~\cite{2022-xilinx-fpga-nvmf}, BlueField~\cite{2022-bluefield}), i10~\cite{2020-nsdi-i10}, ReFlex~\cite{2017-asplos-reflex}\\
Storage + Accelerator & ASIC/CPU~\cite{2016-isca-biscuit,2014-osdi-willow,2017-isca-summarizer}, GPUs~\cite{2012-fast-shredder,2013-asplos-gpufs,2017-tos-spin}, FPGA~\cite{2019-atc-insider,2021-ieee-fermat,2017-vldb-caribou,2020-eurosys-metalfs}, Hayagui~\cite{2020-hotstorage-hayagui}\\ 
%Accelerator + Storage & storage access from accelerators & Gullfoss, FPGA-assisted virtualization\\
Hybrid System & with ARM SoC~\cite{2019-cacm-prog-storage,2022-dfg-cards,2020-asplos-leapio}, BEE3~\cite{2009-msr-bee3}, hybrid CPU-FPGA systems~\cite{2022-asplos-enzian,2019-acm-in-depth-cpu-fgpa}\\ 
\midrule
DPUs with CPU & Fungible (MIPS64 R6 CPU) DPU processor~\cite{2022-dpu-fungible}, Pensando (host-attached P4 Programmable CPU)~\cite{2022-dpu-pensando}, BlueField (host-attached, with ARM CPU)~\cite{2022-bluefield}\\
\midrule
DPUs without CPU & \archName{} (This VIDI proposal, with stand-alone, CPU-free, workload-level NVMe storage \mbox{integration}) \\ 
\midrule  
\end{tabular}
%\vspace{0.1cm}
\caption{Related work in the integration of network, storage, and accelerators devices.}
%\vspace{-1cm}
\label{tab:rwork}
\end{table*}

However, even in this Golden Age, the CPU\footnote{referring to the CPU from the host 
(e.g. x86) as well as smart accelerators like ARM SoC.} remains in the critical path to 
manage data flows~\cite{2022-arvix-bam} (data copying, I/O buffers management~\cite{2020-apsys-p2p-dma}),
accelerators (e.g. PCIe enumeration~\cite{2011-asplos-declarative-pci}), 
and translate between OS-level (packets, request, files) to device-level abstractions (address, 
locations)~\cite{2016-eurosys-posix,2021-hotnets-persistent-packets,2017-systor-flashnet,2021-hotos-zone}). 
Table~\ref{tab:rwork} shows an overview of prior approaches (\Sec{sec:rwork}). 
Additionally, accelerator integration is always done (via virtualization or multiplexing) while keeping 
the CPU and accelerator view of systems resources (DRAM, memory mappings, TLBs) coherent and secure.   
Though necessary, such integration brings complexity to accelerator management and keeps the CPU as the 
final resource arbiter. In contrast to accelerators and I/O devices, the CPU 
performance is not expected to improve by a radical margin~\cite{2021-hotos-last-cpu}, and is even dropping 
with each microarchitectural attack fix~\cite{2020-osdi-cpu-slowdown,2020-cacm-attacks}.
We are not the first one to raise issues associated with the CPU-driven computing 
architecture~\cite{2020-cacm-dsa,2021-hotos-last-cpu}. Despite this awareness, CPU-driven designs and consequently, 
the CPU remains in the critical path of end-to-end system building, thus not escaping the dynamics of 
Amdahl’s Law~\cite{2019-cacm-golden-age}.

The first-principle reasoning suggests the solution: a system where there is no CPU, i.e., a zero-CPU or 
CPU-free architecture.  A completely new computing architecture like zero-CPU will require a radical and destructive 
redesign of computing hardware (buses, interconnects, controllers, DRAM, storage), systems software, and 
applications. A prior example of this approach is the MSR BEE3 system used for emulations~\cite{2009-msr-bee3}. 
A recent example is ETH's Enzian system that designs a hybrid CPU-FPGA dual socket system~\cite{2022-asplos-enzian}. 
The Enzian paper documents the heroic engineering effort it took to design such a system where all board components 
need to be re-designed to integrate an FPGA as a co-processor \textit{with a CPU}. Furthermore, such CPU-centric 
thinking encourages us to inherit and integrate CPU-centric hardware and software choices for an accelerator-centric 
design without re-assessing if such choices make sense and/or can be simplified (see \Sec{sec:case}).

In this work, we take  a more pragmatic approach and investigate the design of a unified NIC-FPGA-storage  
\textit{Data Processing Unit}~(DPU) called \archName{} (Figure~\ref{fig:arch}). \archName{} aims to establish end-to-end hardware 
control/data paths within the DPU without any CPU involvement. The unique design of \archName{} allows us to consider 
building a standalone, self-contained DPU, where no host system is needed to run it, thus reducing the energy cost and 
increasing packaging density. This directly, network-attached FPGA model has 
been used before as well~\cite{2016-ieee-ibm-net-attached-fpga,2014-isca-catapult,2019-nsdi-azure-fpga,2017-vldb-caribou}. 
In this paper, we present a case for such a stand-alone DPU but without a CPU (\Sec{sec:case}), and present design 
choices pertaining to hardware integration (\Sec{sec:dpu-hw}), systems software (\Sec{sec:dpu-sw}), 
and client-interface and workload (\Sec{sec:dpu-client}).  %shows the overall architecture. 

\section{The case for a CPU-free DPU}\label{sec:case}
The CPU-driven design has its clear merits, and its elimination is \textit{not recommended} 
for every workload in general computing. However, for specialized data center workloads (data-parallel,  
accelerator-amenable, disaggregated), the usability of the CPU must be reassessed. There are three primary 
impetuses that encourage us to think about a CPU-free DPU:

First, the era of one-CPU-fits-all is over (design, manufacturing, and thermal 
limits~\cite{2011-isca-dark-silicon,2011-micro-dark-silicon,2011-cacm-future-microprocessor})
and the way forward is \textit{specialization} with reconfigurable hardware and accelerators. The generality of 
the CPU has overheads (i.e., Turing Tax) that hinder specialization for performance or efficiency. 
For example, calculations for the Smith Waterman algorithm in DNA sequence alignment
takes 37 cycles with 40 instructions (35 arithmetic, 15 load/store) with 81 nanoJoules of energy 
(on a 14nm CPU). In comparison, this calculation on a specialized 40nm ASIC takes a single cycle instruction 
with 3.1 picoJoules of energy~\cite{2019-atc-darwin-talk}. The generality and over-engineered design of CPUs for 
any workload also results in poor on-chip resource utilization~\cite{2012-asplos-clearning-cloud}, 
unused silicon~\cite{2011-isca-dark-silicon,2011-micro-dark-silicon}, and elevated security risks~\cite{2020-cacm-attacks}.
At the same time, with the availability of open-source EDA processes and projects~\cite{2022-openroad,2022-chips}, 
exploring workload-specialized hardware designs (with or without CPU) has become more approachable and affordable.

Second, a direct consequence of keeping a CPU-driven design is to inherit its choices of memory addressing, 
translation, and protection mechanisms such as virtual memory, paging, and segmentation~\cite{1970-cacm-virtual-memory}. 
When an accelerator such as FPGA, %\footnote{Using an FPGA as a canonical example for broader non-CPU devices.},  
is attached to a CPU as an external device~\cite{2019-acm-in-depth-cpu-fgpa} or as a 
co-processor~\cite{2022-asplos-enzian}, there is a temptation to provide/port the familiar memory  
abstractions like unified virtual memory~\cite{2020-osdi-os-on-fpga} and/or shared memory~\cite{2020-asplos-optimus}. 
This design necessitates a complex integration with further CPU-attached memory abstractions such as page tables and 
TLBs, virtualization, huge pages, IOMMU, etc., while keeping such an integration coherent with the CPU view of the 
system~\cite{2020-osdi-os-on-fpga,2020-asplos-optimus}. Furthermore, the management of physical memory 
(or the illusion of a flat, uniform physical address space) on modern computing platforms with accelerators and 
heterogeneous CPUs is a non-trivial and complex job~\cite{2020-phd-achermann}. Hence, in this work we argue that 
eschewing CPU and its design baggage, we can explore new memory management designs such as compiler/language-assisted 
solutions even directly on physical addresses~\cite{2022-asplos-caratcake}.

\begin{figure*}[t!]
    \centering
    \begin{subfigure}[t]{0.5\textwidth}
        \centering
        \includegraphics[width=0.8\linewidth]{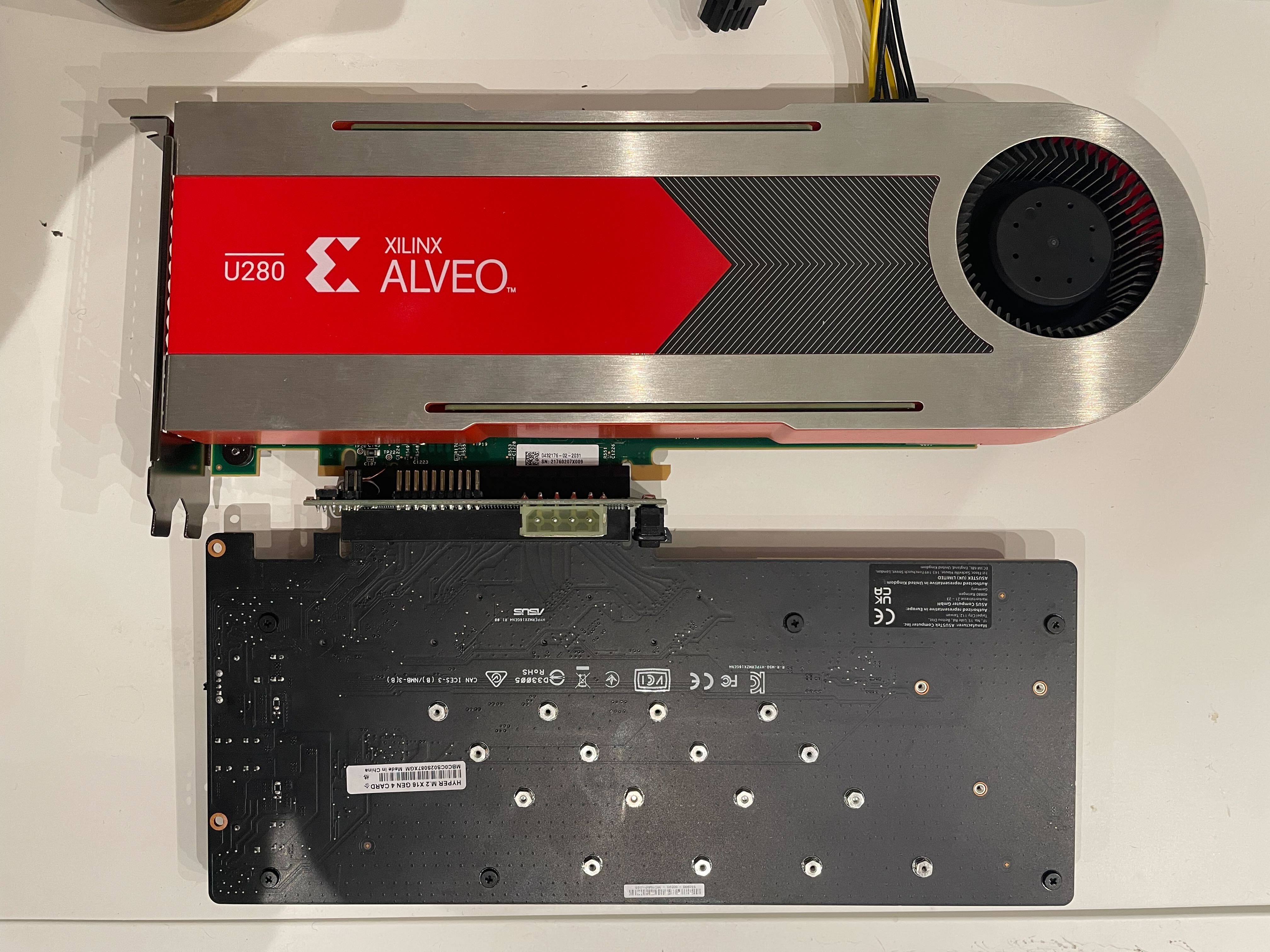}
        \caption{\archName: U280 FPGA-side up.}
        \label{fig:x-front}
    \end{subfigure}%
    ~ 
    \begin{subfigure}[t]{0.5\textwidth}
        \centering
        \includegraphics[width=0.8\linewidth]{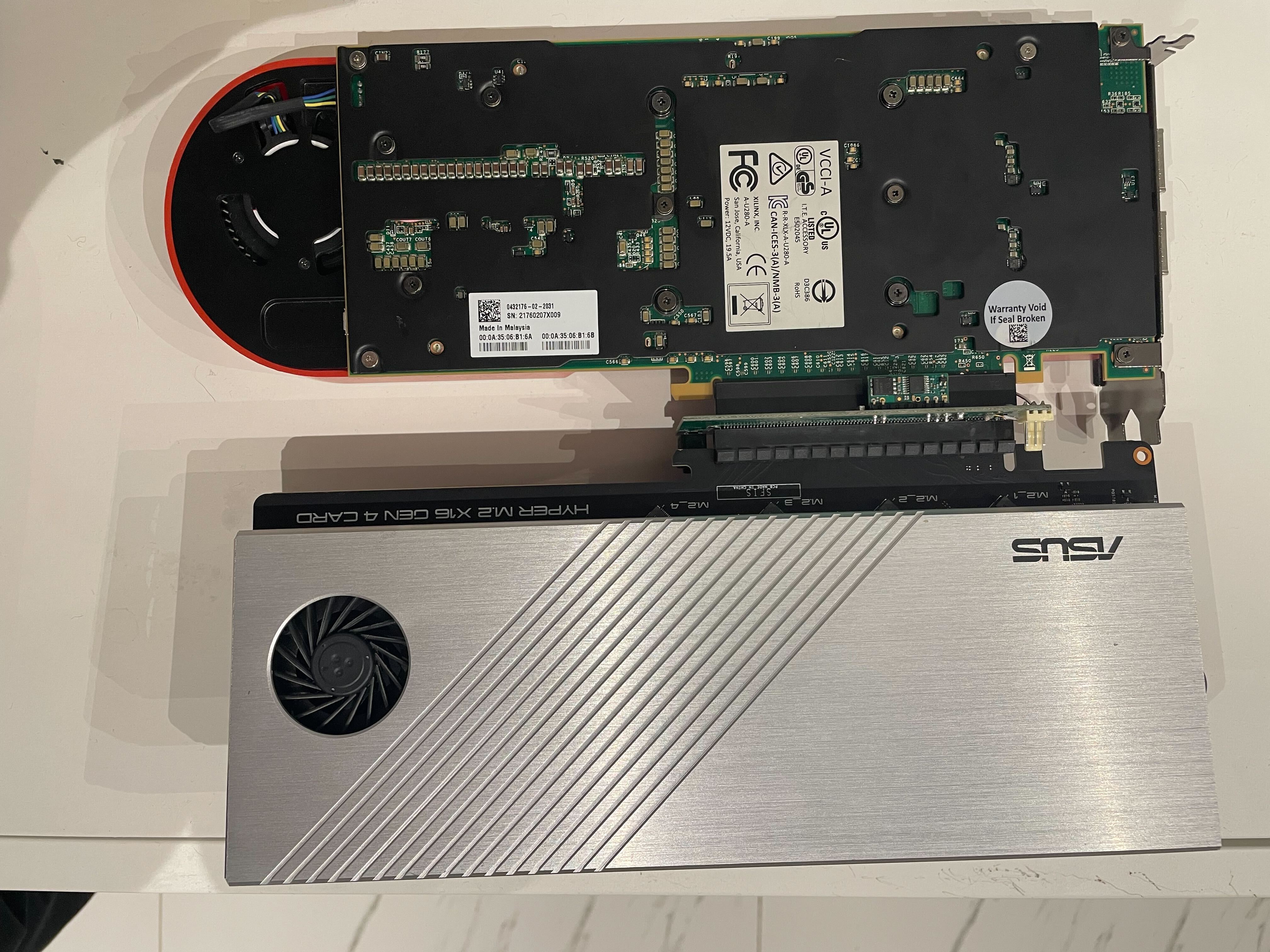}
        \caption{\archName: Asus Hyper M.2 X16 card up with 4x SSDs.}
        \label{fig:x-back}
    \end{subfigure}
    \caption{\archName{} is prototyped with a Xilinx U280 FPGA and 4x NVMe devices with a Asus Hyper M.2 X16 with Samsung SSDs as shown in ~\autoref{fig:x-front} and ~\autoref{fig:x-back}.}
\end{figure*}

Lastly, the CPU-centric design encourages the \textit{active} resources disaggregation where resources 
remain attached to a host CPU that manages the disaggregation logic. This design results in a coarser disaggregation 
granularity with complex and bloated software~\cite{2016-osdi-disaggregation} and a tight integration 
of processor/memory~\cite{2018-osdi-legoos,2022-asplos-clio}. To achieve the vision 
painted by Han et al. in their seminal HotNet'13 paper~\cite{2013-hotnets-disaggregation}, there is a renewed push for 
\textit{passive} disaggregation where disaggregation logic/smartness lies with clients, and a remote resource only 
serves requests as fast as possible~\cite{2018-osdi-legoos,2022-asplos-clio,2020-atc-passive-dissmem,2020-hotcloud-diss-apps,2021-asplos-piberry}.
Passive disaggregation promotes a network-attached model, where memory, storage, DPUs, 
and ASICs are directly connected to a network, and offers a better match with fine-grained computing models like 
Serverless~\cite{2019-cacm-serverless,2019-xxx-serverless-datacenter}. It also enables systems designers 
to rethink 
(i) network protocols for discovery and configuration protocols (e.g., Catapult fabric~\cite{2014-isca-catapult});
(ii) work division between clients and remote servers for distributed resource allocation, and access (e.g., Clio~\cite{2022-asplos-clio}, DUA~\cite{2019-nsdi-azure-fpga}); 
and (iii) offload-friendly abstractions with isolation, multiplexing mechanisms (e.g., group offloading and memory re-assignments~\cite{2019-sigcomm-ipipe,2020-hotcloud-diss-apps,2018-sigcomm-hyperloop}).

\textbf{To summarize:}  In this work, we make a case for the elimination of the 
CPU and its design baggage, and argue that its elimination can bring substantially simplicity 
and offer performance/energy advantages. 
Our attempt to design and implement \archName{} is a step in this direction.

%=====================================================================================================================
%=====================================================================================================================
\section{\archName{}}\label{sec:dpu}
\archName{} is a standalone, network-attached DPU that unifies 100 Gbps Ethernet NIC, FPGA, and NVMe storage 
devices in a single DPU. Figure~\ref{fig:arch} shows the overall architecture of \archName{} with the FPGA board, 
and attached NVMe SSDs. 

\subsection{Hardware Design}\label{sec:dpu-hw}

Commercially, NICs and storage devices (e.g. NVM Express) are available as separate PCIe devices. 
Communication between the two requires control coordination with P2P DMA from the CPU (if supported, e.g., 
NVMe Controller Memory Buffers (CMBs)~\cite{2022-nvme-cmb}) via the PCIe root complex, which typically resides on the 
CPU complex (keeping it in the loop). To make the DPU self-sufficient, \archName{} runs a PCIe root complex with an 
NVMe controller on the FPGA board, which is connected to a 100 Gbps network directly. The FPGA PCIe lanes are connected 
(x16) to off-the-shelf NVMe storage devices via a PCIe bifurcation. Hence, all access to the storage is funneled through 
the FPGA.
With such a design, \archName{} now has an \textit{end-to-end hardware path} from network to FPGA to storage devices. 
The end-to-end hardware path can be \textit{specialized} to a workload with an optimized network transport 
(TCP, UDP, RDMA, HOMA~\cite{2021-atc-homa}), storage API (NVMoF~\cite{2022-xilinx-fpga-nvmf},  
i10~\cite{2020-nsdi-i10}, ReFlex~\cite{2017-asplos-reflex}, KV-SSDs~\cite{2021-systor-kv-ssd})  
with any arbitrary storage functions on the FPGA (compression, pointer chasing, deduplication,  
or application-defined codes).

\textbf{Why FPGA?} Three factors drive the selection of FPGA:
\begin{enumerate}%[leftmargin=10pt,itemindent=0em,nolistsep]
  \item\textbf{Application-specific reconfigurability:}  
The use of FPGA allows us to reconfigure hardware (deep pipelines, unrolled loops, data parallelism, large caches) 
to the best possible implementation of an application-specific logic. ASICs offer similar benefits, but require 
high initial investment and manufacturing costs. Furthermore, as there is an increasing trend to pack 
thousands of workload-specific processing units (PU) in a close vicinity (e.g., 
Cerebras~\cite{2022-cerebras}, Tesla Dojo~\cite{2022-telsa-dojo}), the distance among PUs and memories  
(SRAM, DRAM, or HBM) is of critical importance. Here, we believe that an FPGA-based design offers the best tradeoffs.
 
\item\textbf{Improved FPGA systems software support:} The primary challenge for managing FPGAs comes from carefully 
managing the pipelined execution of the workload with Hardware Description Languages (HDLs).   
With the availability of high-quality DSLs~\cite{2019-asplos-jit-verilog,2018-pldi-spatial,2012-dac-chisel,2018-osdi-amorphos}, 
OS-shells~\cite{2020-osdi-os-on-fpga}, and HDL compilers (hXDP~\cite{2020-osdi-hxdp}), it has become more affordable 
to generate a high-quality HDL for high data rates (100+ Gbps)~\cite{2018-nsdi-azure,2020-osdi-panic}. Overall FPGA 
compilation and debugging processes have also improved~\cite{2022-asplos-fpga-fast-compilation,2022-asplos-fpga-debug}.

\item\textbf{Predictable performance with energy efficiency:} Unlike the CPU and I/O devices that target fine-grained  
time-based statistical multiplexing ($\mu$sec to nsec) to maximize resource utilization, FPGAs target a much 
coarser time-scale (10-100s milliseconds), or even spatial multiplexing which commits resources to a tenant. This 
sharing model helps with building highly \textit{predictable execution pipelines} where once an associated bitstream has  
been sent to the FPGA, the circuit runs a certain clock frequency without any outside interference~\cite{2017-sosp-kvdirect,2016-nsdi-consensus-box}.
The use of FPGAs has been shown to be energy efficient~\cite{2020-osdi-hxdp,2019-atc-insider,2019-ieee-fpga-energy}  
as its energy consumption is proportional to the \textit{active and used} programmable LUTs and the operating 
frequency. Unused logic elements do not consume any energy, resulting in deployments which consume 10-20 Watts, 
which is an order of magnitude less than a server-grade machine~\cite{2016-nsdi-consensus-box}.  

\end{enumerate} 

Apart from the choice of FPGA, \archName{} uses NVM Express (NVMe) for block SSDs, Ethernet for network, and the PCIe 
between FPGA and SSDs. These choices are dictated by practicality and the engineering efforts required. For example, 
the choice of PCIe over other high-performance local interconnects (CXL, CAPI) or networks 
(TrueFabric~\cite{2022-true-fabric}), can be revised as workload demands increase.     

%=====================================================================================================================
%=====================================================================================================================

\subsection{Software and Programming}\label{sec:dpu-sw}
Due to the absence of the CPU and conventional operating system, doing the classical resource 
management with elevated privileges to mediate accesses to a shared resource in \archName{} would be challenging. 
Hence, we must re-negotiate the work division among hardware, compiler, and application with the compiler 
taking a leading role. The role of compiler is not unusual here. It has been shown that compiler-assisted 
designs can help with the traditional OS roles such as for context switching~\cite{2013-tos-compiler-ctx-switch,2018-osdi-dynamic-checkpoints,2021-asplos-compiler-fpga-ctx}, 
memory virtualization~\cite{2022-asplos-caratcake}, single-level memory/storage~\cite{2020-atc-twizzler,2021-asplos-corundum}, 
extraction of parallelism~\cite{2020-osdi-hxdp}, virtualization and multi-tenancy~\cite{2018-osdi-amorphos,2020-asplos-virt-fpga}.

With this compiler-centric approach we run the risk of repeating the failure of the VLIW processors\footnote{VLIW compilers 
were left responsible for parallelism extraction in general workloads, which lead Donald Knuth to  
comment that \textit{``\ldots the "Itanium" approach that was supposed to be so terrific—until it turned out that the wished-for 
compilers were basically impossible to write''}~\cite{2022-knuth-interview}.}. However, we argue that there are two  
fundamental shifts that work in our favor. First, domain/workload-specific architectures are common, and associated 
languages (e.g., OpenCL, Chisel~\cite{2012-dac-chisel}) and compilers are used extensively as the norm. There are 
significant research and commercial interests in co-designing domain- or workload-specific hardware/software. 
Second, unlike VLIW processors, a DPU (specifically FPGA driven) is not aimed to deliver performance for 
all/any workload, hence, restricting the optimization design space. For example, 
hXDP has demonstrated that compile-time heuristics (the Bernstein conditions) with a simple language (eBPF) 
can lend itself to automatic parallelism extraction for packet processing workloads with 
a VLIW softcore processor~\cite{2020-osdi-hxdp}.

Inspired by the LLVM project, in this work we argue that FPGA programming needs to decouple the frontend 
(application logic) and backend (HDL codes) with an accelerator-independent, intermediate representation (IR) 
language. The IR can be used to reason about correctness and safety properties of the program, with compiler-assisted 
transformations for pointer swizzling and privilege calls.       
We make a case that the extended Berkeley Packet Filter (eBPF)~\cite{1993-atc-bpf,2021-ebpf-io} 
language is a suitable match for such an IR for three key reasons. 
First, eBPF is not tied to a specific application-domain and it is used in 
networking~\cite{xdp,2018-conext-xdp}, tracing~\cite{2022-bpf-tracing}, caching~\cite{2021-nsdi-memcache-bpf},  
security~\cite{bpf-seccomp}, and storage~\cite{2019-atc-ebpf-cache,2020-middleware-blockNDP,2020-arxiv-appcode,2021-hotos-ebpf-exo-os}.
It is also supported by healthy, growing communities (Cilium, the eBPF foundation), thus, establishing expertise 
and a knowledge base. Second, due to the simplified nature of the eBPF instruction set, it is possible to verify
and reason about its execution. The Linux kernel already ships with an eBPF verifier~\cite{2018-kernel-bpf-verifier} 
(with simplified symbolic execution checks). Lastly, eBPF supports efficient code generation (via JITing) for multiple 
hardware devices such as x86, ARM, or FPGAs, thus solidifying its position as an accelerator-independent, 
unifying IR for heterogenous computing~\cite{2022-bpf-hetero-uni-abi}. Bear in mind, here we take a broader position 
regarding eBPF where the Linux kernel implementation is one of many possible implementations of an eBPF execution environment. 
For example, there are userspace BPF VMs~\cite{ubpf}, checkers~\cite{2019-pldi-bpf}, and application-specific 
ISA extensions~\cite{2020-osdi-hxdp}. 
Apart from eBPF, we also consider P4, another popular programming language for in-network acceleration (NICs and switches). 
However, P4 programs are designed around packet processing and network abstractions. In restricted capabilities 
(with only filtering and forwarding) there are P4 to eBPF compilers available, though the generality of P4 for 
general data processing is yet to be explored.

\archName{} supports any eBPF-supporting programming language as a frontend. It then uses \texttt{clang/LLVM} 
to generate eBPF IR from the frontend. The eBPF IR is then passed to a two stage compilation process. In the first 
stage, the eBPF IR is passed through the open-source hXDP compiler for parallelism extraction and optimized VLIW   
transformations~\cite{2020-osdi-hxdp,2022-axbryd-hxdp}. In the second stage, the optimized eBPF IR is passed  
through an eBPF-to-HDL compiler for the final HDL code generation. Unlike hXDP, \archName{} runs HDL codes  
directly, not as a VLIW softcore processor on the FPGA.

Beyond the basic compilation of application-provided code to HDL, there are challenges associated with
(i) secure multi-tenant execution; and (ii) FPGA configuration, management, accessibility of data-center 
resources~\cite{2019-nsdi-azure-fpga}. Many past design choices here can be simplified as there are no host system 
resources (on the CPU or OS) that need to be kept coherent and secure while doing execution in the FPGA. We propose 
to leverage the slot-style slicing of FPGA resources~\cite{2018-osdi-amorphos} with a compiler to do workload 
partitioning~\cite{2020-asplos-virt-fpga}. \archName{} runs a configuration kernel that can receive authorized FPGA 
bitstream over the network and assign slices to it.

%=====================================================================================================================
%=====================================================================================================================
\subsection{Client Interface and Workloads}\label{sec:dpu-client}
To provide a client interface that can be \textit{specialized}, \archName{} takes  
% The next big question in the design of \archName{} is the choice of the application interface. 
% Recent trends of passive hardware disaggregation with fine-grained computing~\cite{2019-xxx-serverless-datacenter} allow 
% us to take 
inspiration from Willow~\cite{2014-osdi-willow}, which pioneered an RPC-backed 
programmable SSD interface where a user provides application- and SSD-side RPC stubs. 
Such a flexible design can support any desired specialization of both network as well as storage interfaces. For example, 
we can build network-attached SSDs that can support Corfu consensus protocol~\cite{2012-nsdi-corfu,2013-systor-corfu-hardware}, 
block-level NVMoF accesses, NFS acceleration, or the bump-in-the-wire/near-data execution of application-provided codes 
(B+/LSM tree search, compaction and insertions, file system walks, transactions)~\cite{2019-atc-insider,2020-fast-fpga-lsm}.
Here, we can leverage client-driven request routing~\cite{2014-nsdi-mica} with a shared-nothing, run-to-completion 
datapath~\cite{2014-osdi-ix} for performance.

We focus on three application classes for \archName. First, high volume 
applications such as fail2Ban~\cite{2022-fail2ban}, inspecting and writing network traffic and logs authentication/malicious 
data to attached SSDs. Such applications must handle high volumes of packet data under a tight time budget (100s of 
millions of packets/sec). Second, a latency-sensitive application such as network pointer-chasing. In a disaggregated 
storage, pointer chasing over B+ trees, extent trees, LSM trees (used in many databases, file systems, and key-value 
stores~\cite{2018-cacm-modern-algo-ssd}) results in multiple network RTTs with significant performance 
degradation~\cite{2020-arxiv-appcode}. Lastly, network-attached SSDs that can export application-defined, 
high-level, fault-tolerant abstractions such as trees, lookup-tables~\cite{2021-systor-kv-ssd}, distributed/shared 
logs~\cite{2012-nsdi-corfu,2013-systor-corfu-hardware}, atomic writes~\cite{2011-hpdc-atomic-writes}, 
concurrent appends~\cite{2020-zns-appends}, caches~\cite{2021-nsdi-memcache-bpf}, and concurrent data structures and 
transactional interfaces (similar to Boxwood~\cite{2004-osdi-boxwood}).

One primary challenge here is the composability of multiple functionalities and the state management on FPGA during 
processing. Often storage integration with FPGA is done at the block-level for state-less data processing on data 
streams (such as grep). Hence, appropriate APIs and abstractions are needed to integrate high(er)-level storage abstractions 
with efficient state management on FPGA/BPF such as file systems~\cite{2019-atc-insider,2019-atc-ebpf-cache,2020-eurosys-metalfs},
file/data format integration~\cite{2019-hotstorage-fpga-parquet,2020-ieee-tud-parquet-arrow}, data caching~\cite{2021-nsdi-memcache-bpf},
QoS scheduling (priority sharing of storage/network resources), checkpointing, deduplication, encryption, 
etc. We are in the process of building such modules as shared libraries for FPGA codes.

\subsection{Current Status}\label{sec:dpu-status}
We are prototyping \archName{} with a Xilinx Alveo U280 board which has 2x100 Gbps QSFP~\cite{2022-xilinx-u280}. 
We have designed a PCIe cross overboard~\cite{2022-pcie-xover} to attach 4x NVMe devices to the U280 with 
power\footnote{All \archName{} artifacts (compilers, board design) will be open-sourced.}.

The current system boots in a \textit{stand-alone mode} without any CPU when power is applied and FPGA JTAG self-tests 
are passed. The board is currently attached to a host-system via USB for programming, however, we are in the process 
of developing an OS-shell and control path over the network that can program the FPGA completely independently as well, 
leveraging Partial Dynamic Reconfiguration through the Internal Configuration Access Port (ICAP) of the FPGA.
We have chosen to use a B+ tree key-value store as one of the first applications for \archName{}. 
We have written an XDP-compatible B+ tree that runs on the in-kernel XDP path (in-memory). 
On \archName{}, the tree will store all its data on NVMe devices directly, and will serve get/put/delete requests 
over the network.

Raw latencies of our hardware are: L2 network RTT \textasciitilde$1\mu s$, NVMe latency is $[5-8]\mu s$.
Currently we do not do any caching, hence, all tree access results in an access to the storage device.
With this setup, the average (expected) lookup latencies are: $\mathcal{O}(1 + (tree\_height \times 8) \mu sec))$.
From the past experience with network packet processing pipelines, we expect \archName{} to support $~1$ million 
lookup operations/sec, although the peak performance depends on how many PCIe lanes, NVMe devices, and FPGA kernels 
are running in parallel. 
In our current compilation process, the B+ tree implementation generates 1000+ pipelines stages.
This is one of the largest designs we have tested with our toolchain, which challenges  
the resource availability on the FPGA. Although we are confident that even this unoptimized B+ tree 
implementation can fit on the FPGA and there is plenty of room for optimization to achieve real multi-tenancy.

\section{Related Work}\label{sec:rwork}

Nider and Fedorova also question of the utility of ``the Last CPU'' in the system and investigate the design of 
a \textit{system management bus} to take over the OS/CPU responsibilities~\cite{2021-hotos-last-cpu}. 
Table~\ref{tab:rwork} shows efforts for \textit{pair-wise device} interactions such as 
GPU-with-storage~\cite{2012-fast-shredder,2013-asplos-gpufs,2017-tos-spin,2022-arvix-bam,2022-donard},    
GPU-with-network~\cite{2014-osdi-gpunet,2016-ross-gpurdma,2022-nvidia-gpudirect}, 
accelerator-to/from-storage~\cite{2020-asplos-leapio,2020-hotstorage-acc-sto-nec,2019-atc-m3x,2020-hotstorage-hayagui}, 
SmartNICs~\cite{2019-atc-nica,2020-asplos-lynx,2021-asplos-nic-offloading}, and networked storage 
accesses~\cite{2016-eurosys-flash-disaggregation,2022-xilinx-fpga-nvmf}.
FPGA are explored with (1) networks~\cite{2018-nsdi-azure,2020-osdi-hxdp,2017-sosr-p4fpga,2015-sigcomm-netfpga}; and 
(2) storage~\cite{2019-atc-insider,2014-osdi-willow,2020-eurosys-metalfs}. 
BPF offloading to NIC/FPGA for processing are done with Endance DAG cards~\cite{2021-ebpf-endace},
Netronome~\cite{2021-ebpf-netronome}, Combo6~\cite{2004-xxx-combo6}, but mostly limited to monitoring and traffic 
shaping. 
FPGA-assisted KV stores have considered a close integration of network and KV 
processing (in-memory)~\cite{2013-fpga-mc,2013-hotcloud-fpga-mc-10gbps,2017-vldb-caribou,2017-sosp-kvdirect} and 
selective integration of NAND flash (e.g., BlueDB and Xilinx-KV~\cite{2015-hotstorage-xilinx-kv-40tb-flash,2016-vldb-bluedb}).
One of the closest design inspirations to \archName{} is LeapIO~\cite{2020-asplos-leapio} 
that integrates NVMe flash with RDMA NIC and ARM SoC. \archName{} and LeapIO share the similar motivations  
(cost, energy, and performance efficiency), however, \archName{} could eschew much of design complexity 
of LeapIO (interaction of host x86 CPU and ARM SoC). \archName{} targets a broader design space, where we consider
unification of reconfigurable hardware (here FPGA), network transport 
(100 Gbps Ethernet) and storage (NVMe flash). This unification offers multiple hardware/software specializations  
to support multiple workload needs.

\section{Discussion and Feedback}\label{sec:discussion}
\archName{} is still in its early prototyping phase. From the systems community, we seek feedback on issues like:  

\noindent\textbf{(1) Is eliminating the CPU a worthy pursuit?} In this paper we made a controversial case for removing 
the CPU, and we believe that with the recent hardware and software advancements it is the right time to re-evaluate 
the role of the CPU and the design baggage that it brings. However, we are interested in hearing counter-arguments. 
We understand that beyond technology, operational costs and complexities might put limits to the realization of this 
idea. At what levels of performance, energy, and packaging efficiency gains from a CPU-free design 
will be worth it? The elimination of the CPU-side mediation also necessitates a bigger supporting role from the FPGA 
toolchains, languages, and compilers, a role which was previously split between the host CPU and OS. 
Are FPGA toolchains ready?

\noindent\textbf{(2) What is the right client-interface to build \textit{distributed} \archName{} applications?} 
Looking beyond hardware and a single DPU, what kind of application-level interfaces/abstractions are required for building 
\textit{distributed} CPU-free applications that can be executed over multiple DPUs? A passive resource disaggregation 
puts the responsibility of control coordination on the client-side. Multiple clients either have to coordinate themselves 
or use an external service~\cite{2016-nsdi-consensus-box,2020-osdi-micro-consensus}. However, in order to realize the 
full potential of \archName{}, applications should also reduce the client-side CPU/OS involvement (e.g., use RDMA or DPDK) 
while interacting with \archName. How should one build distributed applications and composable service ecosystems of 
such standalone, passively disaggregated DPUs?

\noindent\textbf{(3) Operational complexity in multi-tenant clouds?} In datacenters, hardware and software fail. 
Tenants are untrusted. The costs of inefficiency and downtime are high. Hence, how to ensure that \archName{} can offer 
secure, multi-tenant execution in FPGAs~\cite{2022-asplos-fpga-enclaves}? How to reduce 
microarchitectural attacks with \archName? Can or should micro-architectural resources of 
\archName{} be managed explicitly with tenants to ensure sufficient isolation with \archName{} DPUs~\cite{2020-hotcloud-stratus}?

\section*{Acknowledgments}
This work is generously supported by the NWO grant number OCENW.XS3.030, \textit{Project Zero: Imagining a Brave CPU-free World!},
and the Xilinx University Donation Program.

\bibliographystyle{plain}
\bibliography{main}

\end{document}